\documentclass[aps,prl, reprint, superscriptaddress]{revtex4-1}
\usepackage{graphicx}

\begin{document}


\title{Observation of migrating transverse Anderson localizations of light in nonlocal media}


\author{Marco Leonetti}
\affiliation{IPCF-CNR c/o Physics Department, University of Rome La Sapienza, P.le A. Moro 2, 00185, Rome, Italy}
\affiliation{Center for Life Nano Science@Sapienza, Istituto Italiano di Tecnologia, Viale Regina Elena, 291 – 00161 Roma (RM) – Italia}
\email[Corresponding Author: ]{marco.leonetti@roma1.infn.it}

\author{Salman Karbasi}%
\affiliation{Department of Electrical Engineering and Computer Science
University of Wisconsin-Milwaukee
Milwaukee, WI 53201, USA}

\author{Arash Mafi}%
\affiliation{Department of Electrical Engineering and Computer Science
University of Wisconsin-Milwaukee
Milwaukee, WI 53201, USA}

\author{Claudio Conti}
\affiliation{Dep. Physics University Sapienza, P.le Aldo Moro 5, I-00185, Roma Italy}

\begin{abstract}
We report the experimental observation of the interaction and attraction of many localized modes in a two dimensional (2D) system realized by a disordered optical fiber supporting transverse Anderson localization.  We show that a nonlocal optically nonlinear response of thermal origin alters the localization length by an amount determined by the optical power and also induces an action at a distance between the localized modes and their spatial migration. Evidence of a collective and strongly interacting regime is given.
\end{abstract}

\maketitle
\textbf{}
In recent years there has been a considerable amount of interest about the direct observation of Anderson localization (AL)\cite{Anderson1958absence, john1987strong, wiersma1997localization, sperling2012direct, gentilini2009ultrashort, levi2011disorder} and the trapping of  waves in a disordered potential. In the field of photonics, various authors have reported evidences of the transition to the Anderson regime  and of the interplay between localization and nonlinearity. The \emph{transverse} localization\cite{schwartz2007transport, szameit2010observation} due to disorder sustains non-diffracting beams in media in which the refractive index is randomly modulated orthogonally to direction of propagation.  Recent experimental results stimulated a large body of theoretical work dealing with the role of self-focusing and defocusing in the evolution of the disorder induced localizations.

There is a relevant debate about the fact that nonlinearity\cite{lahini2008anderson,muskens2012, rose2012nonlinear,sapienza2010cavity, rose2012nonlinear} may enhance, or hamper, this linear trapping mechanism, and there are many open research directions, as for example considering quadratic nonlinearities \cite{Flach2009Universal, folli2012anderson, Conti2012Solitonization, Folli:13, Jovic:13}.
Indeed the interplay between disorder and a nonlinear response is expected to alter the commonly accepted scenario about the absence of diffusion and transport in the Anderson regime, when all the states are exponentially localized. Particularly intriguing is the role of spatial nonlocality. Indeed, if disorder induces exponential localizations and reduces the interactions of distant modes, nonlocality (i.e., a nonlinear perturbation that extends far beyond the region of interaction) is expected to create some action at a distance dependent on power.

\begin{figure}
\includegraphics[width=\columnwidth]{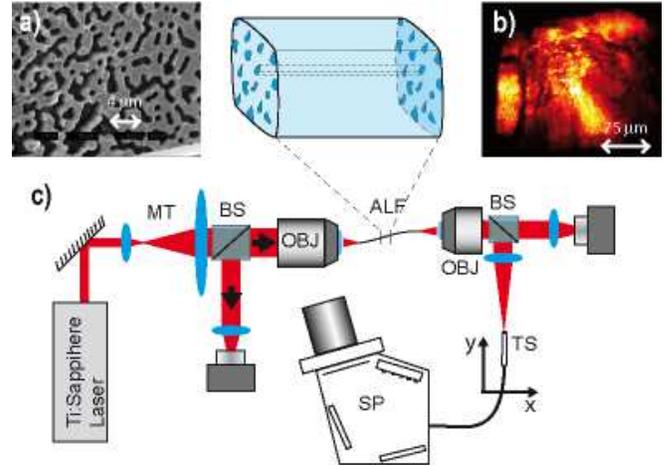}
\caption{ (a) Scanning electron microscope image of the fiber facet; (b) image of the fiber exit face when the input is completely illuminated; (c) sketch of the experimental setup.  \label{fig:figure1}}
\end{figure}

Despite some theoretical investigations\cite{folli2012anderson}, AL with highly nonlocal nonlinearity was never experimentally investigated before; here we report on the experimental investigation of light localization in a 2D disordered fiber with a nonlinearity of thermal origin, which is known to be highly nonlocal\cite{turitsyn1985spatial, Krolikowski2002Collapse, ContiPRL04, rotschild2005solitons, Folli12, Belic2013, Desya2013}. We show that many exponentially localized states can be simultaneously excited by a broadband laser beam. We measure their localization length in terms of the optical power, and give evidence of the action at a distance between localized states. We also find that, because of nonlocality, the AL migrate from their position and move in a collective way. These experimental findings are quantitatively explained by a theoretical approach based on the highly nonlocal approximation (HNA).

We investigate the nonlinear propagation of light in fibers with a binary distribution of the refractive index. The disordered fibers under study are fabricated as described in \cite{karbasi2012observation} by realizing a binary matrix of Polymethyl-methacrylate (PMMA, refractive index $n_{PMMA}=1.49$), and Polystyrene (PS, $n_{PS}=1.59$) resulting in an index contrast $\Delta n$ =0.1 (an image of the fiber tip is reported in figure \ref{fig:figure1}a) which is three orders of magnitude stronger than previous investigations \cite{schwartz2007transport}. The complete characterization of the linear Anderson localization in our samples (including averaging over the disorder) has been given in references \cite{karbasi2012observation, karbasi2012detailed}, while calculations \cite{de1989transverse} predict a transverse localization length two orders of magnitude smaller than the fiber size.

We measure the localization length of transverse Anderson modes (AM) observed when injecting in the fiber a broadband laser (Ti:Shappire oscillator model Coherent Mantis; 500 mW power, 80 nm bandwidth, 820 nm center wavelength, 15 fs pulse duration) through a long working-distance microscopy objective, which produces a spot size between 0.8 and 10 $\mu m$ at the entrance of the fiber. As described in figure \ref{fig:figure1}, the output face of the fiber is imaged both on a CCD camera and on a fiber coupled to a spectrograph, which retrieves the output emission from an area with sub-micrometer spatial extension. Two computer-controlled motors allow to vary the fiber position in the virtual image plane with a spatial resolution below one micron, and enable to measure the spatial distribution of Anderson localizations at different wavelengths.

By the CCD camera, we observe at the fiber output a single bright spot located in correspondence of the input beam position at the entrance. For an input power of the order of $1mW$, the average measured localization length is 45 $\pm$ 5 $\mu m$\cite{karbasi2012detailed}. When increasing the input power to 35 mW, the shape of the output beam gets sharper and localization length lowers  to 30 $\mu m$. This self-focusing action cannot originate by the intrinsic material nonlinearity because the plastic materials sustain only a thermal de-focusing nonlinearity.  This optical nonlinear response is found to occur on a timescale of the order of one second; this indicates that the contraction of the spot size is due to thermal nonlinearity (PPMA light absorption is 2500 $ dB/km$, PS absorption is 600 $ dB/km$, PMMA thermal conductance is $K_{PMMA}=0.25 W/mC^\circ$, and PS thermal conductance $K_{PS}=0.033 W/mC^\circ$), which is temporally slow and inherently nonlocal\cite{rotschild2005solitons, turitsyn1985spatial}. Indeed, even if PS and PMMA show a nearly identical dependence of the refractive index with temperature, PS absorption is about four times smaller than PMMA,  and PMMA thermal conductivity is one order of magnitude larger than PS. This results in the presence of temperature hotspots  located in the PMMA, which induce a local decrease of the refractive index and thus an increase of the refractive index mismatch. By numerically solving the heat transfer equation we found that an input power of 10 mW produces temperature gradients that increase the refractive index mismatch of an amount of the order of $\Delta n_{NL}=10^{-5}$. For comparison we performed optical measurements for a homogeneous (without disorder) fiber. We retrieved a localization length of the order of the transverse fiber size (hundreds of microns) and not affected by the power level.

\begin{figure}
\includegraphics[width=\columnwidth]{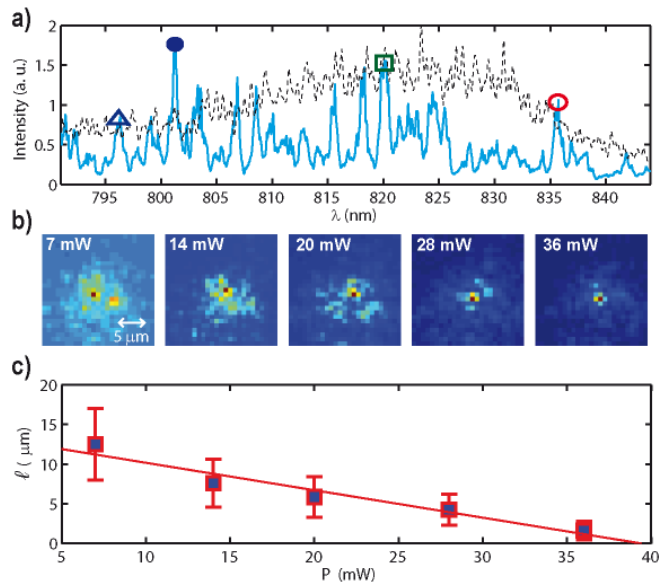}
\caption{\label{fig:figure3}  (a) output transmitted spectrum from the disordered fiber;  the input spectrum is shown as a dashed line; (b) spatial distribution for the most intense mode  of (a) indicated by the full circle, versus the input power $P$; (c) average localization length $l$ estimated by Gaussian fit (error bars are obtained from the statistics over 35 modes). Fit results: $l(0)=14~\mu$m, $P_C=12.5$~mW; fiber length z=8cm.}
\end{figure}

A striking  feature is that the spectrum retrieved at the output is strongly differing from the input spectrum: it is composed by sharp peaks appearing at various frequencies as shown in figure \ref{fig:figure3}a. These peaks demonstrate multi-color transverse localization of light, and allow to follow the behavior of any mode as a function of the input optical power.
The image of the mode indicated by a full circle in figure \ref{fig:figure3}a is reported in the panels of \ref{fig:figure3}b. We measure the average mode size (estimated by a Gaussian fit of the intensity profile) for a set of 35 modes and obtain a linear scaling with the pump power (see figure \ref{fig:figure3}c). The origin of this behavior is the following: at low power the laser light increases the refractive index mismatch between PS and PMMA, strengthening the localization, which arrives to confine light down to an area comparable with the size of the wavelength when input power is 35 mW. A direct comparison of the localization length on different disorder realization (different fibers) in the nonlinear regime is tricky due to variations in confinement and losses across different samples. However we found a linear decay of the modes-averaged localization length $\ell$ in all the fibers we studied.

We remark that two competing phenomena are present: (i) a focusing effect due to the interaction between nonlinearity and randomness, and driven by the increasing refractive index mismatch that enhances the strength of disorder ($l$ decreases when increasing $\Delta$n \cite{folli2012anderson,Conti2012Solitonization}), and (ii) a thermal de-focusing nonlinearity for the bulk PS and PMMA. Hence we have evidence that the interplay between the binary structure and the Anderson localization turns a de-focusing nonlinear response into a focusing one, producing a spatial compression of the localization down to the diffraction limit \cite{leonetti:051104}.

The variation of $l$ with the nonlinear nonlocal response can be theoretically described by the highly nonlocal approximation (HNA) \cite{Snyder97}. Specifically, we model the system by the paraxial wave propagation
\begin{equation}
2ik \frac{\partial A}{\partial z}+ \nabla^2_{x,y} A+ 2 k^2 \frac{\Delta n}{n_0} A=0\text{,}
\label{paraxial1}
\end{equation}
where $A$ is the optical field normalized such that $|A|^2$ is the intensity, $k=2\pi n_0/\lambda$ is the wavenumber with $n_0=(n_{PS}+n_{PMMA})/2$ the average refractive index.
$\Delta n=n_{PS}-n_{PMMA}=\Delta n_R+\Delta n_{NL}$ is the index perturbation including two contributions: (i) a linear perturbation $\Delta n_R(x,y)$ due to the disorder;
and (ii) a nonlinear and nonlocal term, dependent of $|A|^2$, which is written as (see, e.g., \cite{turitsyn1985spatial, Krolikowski2002Collapse})
\begin{equation}
\Delta n_{NL}=\int K(x-x',y-y') |A|^2(x',y') dx' dy'\text{.}
\label{nonlocalsusc}
\end{equation}
In Eq.(\ref{nonlocalsusc}), $K(x,y)$ is the Kernel function which, for a thermal nonlinearity, is given by the Green function
of the Fourier heat equation, which, in general cannot be written explicitly for the considered geometry.
The paraxial approximation in (\ref{nonlocalsusc}) can be removed by resorting to the Helmholtz equation, which is
not done here for the sake of simplicity.

As the temperature varies on a spatial-scale comparable
with the fiber size much larger than the wave localization length,  a specific localization $|A|^2$ is much more localized than $K(x,y)$,
so that the latter can be taken out of the integral in (\ref{nonlocalsusc}), letting $r^2\equiv x^2+y^2$ we have
\begin{equation}
\Delta n_{NL}\cong K(x,y)\int |A|^2 d\mathbf{r}\cong P \left(\Delta n_1 +\frac{r^2}{2} \Delta n_2\right)\text{.}
\label{nonlocalsusc2}
\end{equation}
In (\ref{nonlocalsusc2}) $P$ is the power of the single localized state, and we
expanded $K(x,y)$ in a Taylor series centered on the AL location $r=0$. As we are dealing with exponentially localized states, we can further neglect, for the moment, the term weighted $\Delta n_2$, and treat the Kernel function as a constant: $K(x,y)\cong K(0,0)\equiv \Delta n_1$, being $\Delta n_1$ the nonlinear index perturbation at a unitary input power $P$. Eq.~(\ref{nonlocalsusc2}) shows that the leading effect of the nonlocal nonlinearity $\Delta n_{NL}$ is modifying the refracting index perturbation $\Delta n_R$ and hence shifting the eigenvalues of the the disorder induced localized states. In the HNA, these are given by the solutions $A(x,y,z)=a(x,y)\exp(i\beta z)$ with \begin{equation}
-\frac{1}{2k}\nabla^2_{\perp} a- \frac{k}{n_0}\Delta n_R(x,y) a=
\left(\beta+k\frac{\Delta n_1 P}{n_0}\right) a \text{.}
\label{boundstates}
\end{equation}

In the absence of nonlinearity $\Delta n_1=0$ in (\ref{boundstates}), and the ground state is exponentially localized with negative eigenvalue $\beta=\beta_L<0$. The effect of nonlinearity is to shift the eigenvalue in a power dependent way, so that we have $\beta(P)+k \Delta n_1 P/n_0=\beta_L$.
It is important to stress that in the case under consideration, because of the different thermal conductivities of PS and PMMA, the effect of the increased temperature is to enhance the refractive index modulation and hence the strength of disorder, so that $\Delta n_1>0$; this implies that the negative eigenvalue $\beta(P)=\beta_L-k\Delta n_1 P/n_0$ decreases with $P$. As the localization length $L$ scales as $1/\sqrt{2k|\beta(P)|}$, \cite{LifshitsBook} this induces the self-focusing action observed in the experiments thus producing a behavior $l(P)=l(0)/\sqrt{1+P/P_C}$ with $P_C=n_0 |\beta_L|/2 k\Delta n_1$. In the experiments shown in Fig.\ref{fig:figure3}c, we fit data by a linear approximation of the previous function: $l(P)=l(0)(1-\frac{P}{2P_C})$.
From the experimental data we find $l(0)\cong 1/\sqrt{2k|\beta_L|}\cong 14~\mu$m and $P_C\cong 12.5$~mW, which give $\Delta n_1\cong 10^{-3} W^{-1}$, so that at power $P=10~$mW we estimate $\Delta n_{NL}=\Delta n_1 P=10^{-5}$, in agreement with the estimation from numerical solution of the heat transfer equation.

\begin{figure}
\includegraphics[width=\columnwidth]{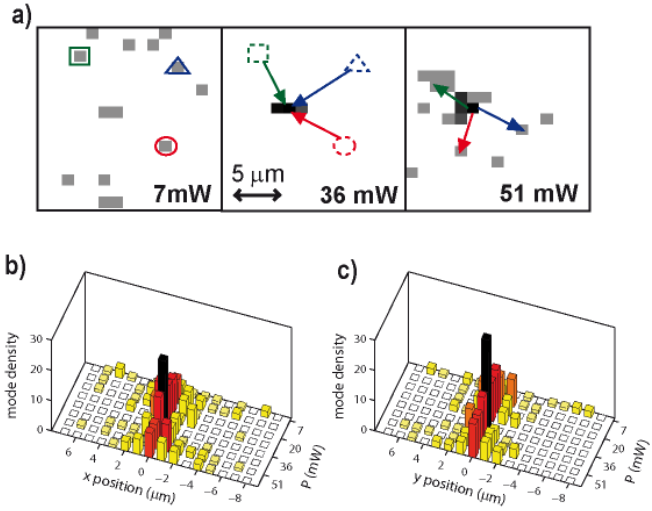}
\caption{\label{fig:figure4}  (a)
position of several Anderson states at different frequencies at different powers showing the nonlinearity driven motion. Some modes are highlighted by geometrical shapes corresponding to spectral peaks
in panel \ref{fig:figure3}a, arrows indicate their motion upon increasing power; (b) mode density as a function of power and of the $x$ position;(c) and in (b) for the y position. The localizations first migrate towards the same position, and then, after colliding, spread again.}
\end{figure}

We next show evidence of an even more intriguing effect, related to the ALs positions. Figure \ref{fig:figure3}b shows that at the maximum degree of localization, most of the intensity of a single mode is collapsed in a single pixel, which gives the AL location. By using this information we retrieve the mode-density  $\rho$ (number of modes per pixel) shown in figure \ref{fig:figure4}a; darker pixels represent positions in which many modes at different frequencies are simultaneously present. The modes position of the 35 most intense peaks of the spectrum is approximated with their intensity maximum, then the mode-density is retrieved counting the number of modes present per pixel: when a single pixel (our experimental spatial resolution) at position (x,y) hosts many modes, $\rho(x,y)$ is high; on the contrary an empty pixel at (x,y) corresponds to a vanishing $\rho$.

At low power, ALs are sparsely distributed in space; when increasing the power, they ``migrate'' towards the peak of optical intensity. Modes translate of an amount greater than 12 $\mu m$, i.e., a distance one order of magnitude larger than their spatial extension, when power is increased from $7$mW to $36$mW. The same phenomenon is represented in the three dimensional plots in panels \ref{fig:figure4}b, c. Mode density  is reported on the z-axis in terms of the transverse  position  $x$ ($y$ position in panel \ref{fig:figure4}c) and of the injected power. All modes collapse in the same position when the input power is about 35 mW, while they spread again at higher power.

\begin{figure}
\includegraphics[width=\columnwidth]{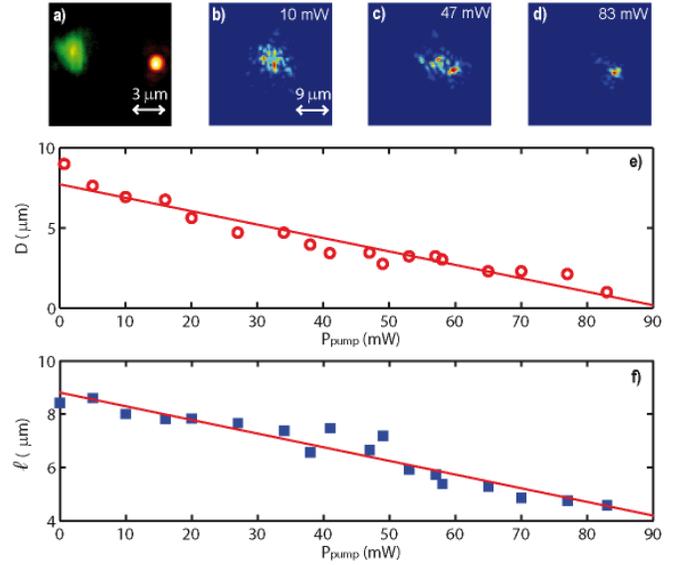}
\caption{\label{fig:figure5} (a) image of the entrance of the fiber (scale is 3$\mu m$); (b,c,d) fiber exit corresponding, respectively,  to 10, 47 and 83 mW input power, the pump beam has been eliminated by a laser line filter (scale is 9 $\mu m$); (e) displacement D versus $P_{Pump}$ , fitted with equation 8 (D(0)= 7 $\mu$m; $\frac{|\Delta n_2|z^2}{2n_0}$=0.01); (f) localization length $\ell$ as a function of $P_{pump}$ and fit as in Fig. \ref{fig:figure3}c.}
\end{figure}

To show that this collective motion is driven by a nonlocal action at a distance between localized states, we designed an experiment with only two excited modes.
We inject in the fiber two beams: (i) a ``probe,'' low power $P_{probe}$ green laser, at wavelength 532 nm   ( continuous wave Nd:YAG  Ventus Laser Quantum, 500 mW maximum power),
and (ii) a ``pump,'' broadband high-power $P_{pump}$ infrared (wavelength 800 nm) laser.
Light reflected from the entrance of the fiber is shown in figure \ref{fig:figure5}a, and allows to measure the distance between the two input spots, which is about 9 $\mu m$.
Both the beams excite ALs. Panels \ref{fig:figure5}b,c,d  show the probe light (a laser line filter in front of the camera allows to collect only light at 532 nm) at the exit of the fiber for various powers of the pump beam (at 800 nm).
We observe two effects: (i) the controlled steering of the Anderson localization at $532$nm towards the position of the control beam (the ``migration''), (ii) a contraction of its spatial extension, that is an all-optically controlled localization length. These features are quantified in the panel \ref{fig:figure5}e, which shows the displacement of the probe position $D$ (open circles) and in panel \ref{fig:figure5}f, which reports the localization length $l$ ( full squares ) as a function of the power in the pump beam.

To support the previous findings by a theoretical ana\-lysis we resort to an approach originally developed for solitons \cite{Conti05Complex}. Specifically, we denote the mean position of any of the exponential localizations by a two dimensional vector $\mathbf{r}_p=(x_p,y_p)$, with $p=1,2,...,N$ and $N$ the number of localizations. By writing the optical field $A$ as a superposition of ALs, and by using the Ehrenfest theorem of standard quantum mechanics on Eq.(\ref{paraxial1}), following the arguments in \cite{Conti05Complex}, it turns out that in the presence of a nonlinear perturbation $\Delta n_{NL}$, the following equations holds true ($\mathbf{r}=(x,y)$).

\begin{equation}
P_p \frac{d^2 \mathbf{r}_p}{d z^2}=\int I_p(\mathbf{r}-\mathbf{r}_p) \nabla_{x,y} \frac{\Delta n_{NL}}{n} d\mathbf{r}\text{,}
\label{newton1}
\end{equation}
which describe the motion in the $z-$direction of the states under the action of the nonlinearity. In (\ref{newton1}), $I_p(\mathbf{r}-\mathbf{r}_p)$ is the intensity profile of the AL with index $p$ and power $P_p=\int I_p(\mathbf{r}-\mathbf{r}_p) d\mathbf{r}$. As the various localizations do have different wavelengths, they are incoherent, and $\Delta n_{NL}$ can be written as the sum of their respective index perturbations $\Delta n_{NL,q}(\mathbf{r}-\mathbf{r}_q)$ with:
\begin{equation}
\Delta n_{NL}=\sum_{q=1}^N \Delta n_{NL,q}\cong \sum_{q=1}^N \frac{P_q \Delta n_2}{2} (\mathbf{r}-\mathbf{r}_q)^2\text{.}
\label{deltanall}
\end{equation}
In (\ref{deltanall}) after the HNA as in Eq.(\ref{nonlocalsusc2}), we expand $\Delta n_{NL,q}(\mathbf{r}-\mathbf{r}_q)$ as a Taylor series with respect to the spatial coordinate centered in the localization positions $\mathbf{r_q}$, and omitted the constant term $\Delta n_1$ as it does not affect the change in the AL location because of the operator $\nabla_{x,y}$ in (\ref{newton1}). At variance with Eq.(\ref{nonlocalsusc2}),
we retained the higher order term weighted by $\Delta n_2<0$, which induces the action at a distance between localizations.
In other words, $\Delta n_2$ is negligible when considering the localization length of the single state, but sustains the interactions of distant AL.
By using (\ref{deltanall}) in (\ref{newton1}) and treating $I_p$ as a Dirac $\delta$ with area $P_p$,
because of the strong localization of the AL with respect to the refractive index profile, we have
\begin{equation}
P_p \frac{d^2 \mathbf{r}_p}{d z^2}=-\nabla_{x_p,y_p} \sum_{q=1}^N \frac{|\Delta n_2|P_q P_p}{2 n_0} | \mathbf{r}_p-\mathbf{r}_q   |^2\text{.}
\label{newton2}
\end{equation}
Equation (\ref{newton2}) predicts that the ensemble of disorder induced localizations behaves as a system of interacting particles with pairwise attractive potential \cite{Conti05Complex}, and the conservative force between two states is proportional to the product of their two powers, as in a gravitating system the attractive force is proportional to the product of masses according to the Newton law.
As a result of this analogy, the more powerful localizations will attract the others, and the whole system will tend to collapse in a specific point, as observed in our experiments. After the collapse and the interaction, being the system conservative, the kinetic energy will convert into potential energy, and
the localization will spread again (see Fig.\ref{fig:figure4}a).

In the simplest case of only two localizations, with powers $P_{probe}$ and $P_{pump}>>P_{probe}$, the probe localization (green beam in figure \ref{fig:figure5}), has a displacement $D(z=0)$ at the fiber input from the pump beam. By Eq.(\ref{newton2}), after a propagation distance $z$,
at the lowest order in $|\Delta n_2|$, the displacement is reduced and given by

\begin{equation}
D(z)=D(0)\left(1-\frac{|\Delta n_2| z^2}{2 n_0} P_{pump}\right)\text{.}
\label{displacement}
\end{equation}
Eq.(\ref{displacement}) shows that the probe position scales linearly with the pump beam power $P_{pump}$, as shown by a best fit in Fig.\ref{fig:figure5}e.

In summary, we have demonstrated a novel form of self-focusing action occurring in a disordered fiber with a binary index distribution and relying on the de-focusing thermal nonlinearity; this effect is due to the nonhomogeneous temperature distribution that increases the refractive index mismatch and strengthen the transverse Anderson localization. This broadband nonlocal phenomenon generates a migration of disorder-induced localized modes, a form of transport achievable in the Anderson regime. The possibility of steering the position of the Anderson states and control transmission channels in a random material may sustain potential applications and is an evidence of the complex regimes achievable in disordered nonlinear optical propagation.



%

\section{Acknowledgements}     S.K. and A.M. are supported by grant number 1029547 from the National Science Foundation.

\end{document}